An examination of demographic differences in obtaining investment and financial

planning information

by

Paul L Bechly

A dissertation submitted to the faculty of

Wilmington University in partial fulfillment

of the requirements for the degree of

Doctor of Business Administration

Wilmington University

Spring 2018

An examination of demographic differences in obtaining investment and financial
planning information

by

Paul L Bechly

I certify that I have read this dissertation and that in my opinion it meets the academic and
professional standards required by Wilmington University as a dissertation for the degree of
Doctor of Business Administration.

Signed:_______________________________________________________________

     W. Daniel Young, Ph.D., Chairperson of Dissertation Committee

Signed:_______________________________________________________________

     John L. Sparco, Ph.D., Member of the Dissertation Committee

Signed:_______________________________________________________________

     Guillermina Gonzalez, DBA, Member of the Dissertation Committee

Signed:_______________________________________________________________

     Kathy S. Kennedy Ratajack, DBA, Dean, College of Business

Acknowledgements

To my dissertation committee; Dr. Young, Dr. Sparco, and Dr. Gonzalez; whose diligence and expertise contributed greatly to the success of this project.

To the Doctor of Business Administration program professors at Wilmington University who challenged me to learn and grow from an academic perspective.

To my peers in DBA Cohort 14, for which we had the opportunity to learn together.

To my family and friends who understood the time commitment associated with this program; and provided motivating encouragement to stay focused through completion.

To my professional support team at Morgan Stanley; H. Rodney Scott, John Caven, Chris Connolly, Vince Crognale, Edward Duffy, and Steven Brettler; all of whom either supported this Doctoral program directly or supported my professional growth as managers during my 25-year career as a financial advisor.

To the leadership of Wilmington University who established the Doctor of Business Administration program ten years ago, and subsequently developed the program into the accredited and nationally respected program that it has become today.

# Table of Contents

## List of Tables

# Abstract

Financial literacy and financial education are important components of modern life. The importance of financial literacy is increasing for financial consumers because of the weakening of both government and employer-based retirement systems. Unfortunately, empirical research shows that financial consumers are not fully informed and are not able to make proper choices even when appropriate information is available. More research is needed as to how financial consumers obtain investment and financial planning information. A primary data study was conducted to understand the differences between the demographic categories of gender, age, education-level, and income-level with the means of obtaining investment and financial planning information. In this research study, which selected a population from the LinkedIn platform, statistical differences between gender, age, education-level, and income-level were confirmed. These differences helped to confirm prior research in this field of study. Practical opportunities for commercial outreach to specific populations became evident through this type of research. Providers of investment and financial planning information can access their targeted audience more effectively by understanding the demographic profile of the audience, as well as the propensity of the demographic profile of the audience to respond. As this type of research is relatively easy to construct and administer, commercial outreach for providers of investment and financial planning information can be conducted in a cost-efficient and effective manner.

# CHAPTER 1

# INTRODUCTION

## Background

Financial literacy and financial education are important components of modern life. Nearly every adult in the United States, and also the world, needs at least a basic level of financial literacy to be able to function in society (Johnson & Lamdin, 2015).

The level of financial literacy is not satisfactory in most countries, and the most important means to improve financial literacy is financial education (Fraczek, 2014). Financial education is the process by which financial consumers or investors gain understanding of financial products, concepts, and risks (Lusardi, 2015a). By obtaining information, instruction, and advice, financial consumers develop the skills necessary to be able to make informed choices (Fraczek, 2014). The importance of financial education has increased globally because of the increasing transfer of responsibilities to individuals, and away from traditional entities such as employing corporations and governments (Fraczek, 2014). As such, financial consumers are becoming increasingly responsible for their own financial wellbeing (Johnson & Lamdin, 2015).

The global economy requires financial literacy of its citizens. People need to understand the basic concepts of value and monetary exchange for products and services (Fraczek, 2014). As people gain more monetary resources than what is needed for daily living, education is necessary to plan for strategic goals such as housing, capital goods purchases, business ownership, and retirement income (Johnson & Lamdin, 2015). Some financial literacy topics, such as personal budgeting and time value of money, are relatively simple to learn (Lusardi, 2012). Financial education is beneficial for more complex financial literacy tasks such as investing and managing debt (Fernandes, Lynch, & Netemeyer, 2014). Some financial literacy topics such as retirement planning, tax planning, insurance planning, and estate planning are highly complex and subject to continual change (Johnson & Lamdin, 2015). For these advanced financial literacy topics, in addition to the financial education necessary to learn these concepts, additional ongoing financial education is needed to maintain a level of financial literacy that remains current with changing rules and conditions (Lusardi & Mitchell, 2007b).

Structured ongoing financial education is not always conveniently available for financial consumers, but fortunately, financial consumers are information seekers and there is an abundance of media to meet their needs (Mezick, 2002). Unfortunately, financial consumers are faced with too many choices for financial products and services (Harrison, 2003). Financial services providers understand this and recognize the role of media as a tool for enhancing public confidence (Stankovska & Popovski, 2012)

## Problem Statement

The importance of financial literacy is increasing for financial consumers as a result of the weakening of both government and employer-based retirement systems (Xiao, Chen, & Chen, 2014). The Social Security system, which is the traditional retirement income source in the United States, has become underfunded (Xiao, et al., 2014). Private-sector pension systems are changing from defined-benefit plans, where the employer is responsible for performance, to defined-contribution plans, where the employee is responsible for performance. As such, financial consumers need to be concerned about their long-term economic security (Xiao, et al., 2014).

Unfortunately, empirical research shows that financial consumers are not fully informed and are not able to make proper choices even when appropriate information is available (De Mesa, Irlenbusch, & Reyniers, 2008). Moreover, to complicate this situation, financial consumers are exposed to an increasing assortment of financial products and services (Harrison, 2003). Financial consumers must also choose between a multitude of media offerings as sources for investment and financial planning information (Mezick, 2002). The challenges of personal financial planning and the universe of media sources for reliable information is creating an environment of uncertainty and anxiety for financial consumers (Harrison, 2003).

## Significance of Study

This study has significance related to the magnitude of the problem statement, which extends to nearly the entire population of the United States (Johnson & Lamdin, 2015). Furthermore, socio-demographic factors have influence over the level of financial literacy for financial consumers (Fraczek, 2014). Key demographics for research include gender, age, and life situation demographics such as education level and income level (Fraczek, 2014). Research conclusions may provide answers related to questions involving financial awareness, knowledge, skills, and even attitudes of financial consumers and investors (Fraczek, 2014). Empirical studies indicate that individuals who seek investment and financial planning information are more successful in achieving their financial goals over time (Lusardi & Mitchell, 2007b).

More research is needed as to how financial consumers actually obtain investment and financial planning information (Harrison, 2003). There needs to be greater understanding of where financial consumers actually source this information, and if the information is helpful in increasing financial literacy (Harrison, 2003). It is also important to understand how different demographic groups make use of media to obtain investment and financial planning information. The insight provided by such research will be potentially valuable to financial service providers, government entities, social media providers, academic researchers, and to the investing public in general (Stankovska & Popovski, 2012).

## Research Question

This study seeks to answer the following research question that evaluates how financial consumers obtain investment and financial planning information based upon the four key demographics of gender, age, education-level, and income-level:

**RQ1.   How are gender, age, education-level, and income-level associated with independent use of the Internet, versus the use of professional advisors, for obtaining investment and financial planning information?**

## Hypotheses

The sixteen hypotheses stated below are intended to identify differences in information sourcing by financial consumers based upon the four key demographics of gender, age, education-level, and income-level:

**H1a.    There is a statistically significant difference in use of the Internet to gather investment and financial planning information based upon gender.**

**H1b.    There is a statistically significant difference in frequency of Internet use to gather investment and financial planning information based upon gender.**

**H1c.    There is a statistically significant difference in use of professionals to gather investment and financial planning information based upon gender.**

**H1d.    There is a statistically significant difference in frequency of professional use to gather investment and financial planning information based upon gender.**

**H2a.    There is a statistically significant difference in use of the Internet to gather investment and financial planning information based upon age.**

**H2b.** There is a statistically significant difference in frequency of Internet use to gather investment and financial planning information based upon age.

**H2c.** There is a statistically significant difference in use of professionals to gather investment and financial planning information based upon age.

**H2d.** There is a statistically significant difference in frequency of professional use to gather investment and financial planning information based upon age.

**H3a.** There is a statistically significant difference in use of the Internet to gather investment and financial planning information based upon education level.

**H3b.** There is a statistically significant difference in frequency of Internet use to gather investment and financial planning information based upon education level.

**H3c.** There is a statistically significant difference in use of professionals to gather investment and financial planning information based upon education level.

**H3d.** There is a statistically significant difference in frequency of professional use to gather investment and financial planning information based upon education level.

**H4a.** There is a statistically significant difference in use of the Internet to gather investment and financial planning information based upon income level.

**H4b.** There is a statistically significant difference in frequency of Internet use to gather investment and financial planning information based upon income level.

**H4c.** There is a statistically significant difference in use of professionals to gather investment and financial planning information based upon income level.

**H4d.    There is a statistically significant difference in frequency of professional use to gather investment and financial planning information based upon income level.**

<div align="center">

**Definition of Terms**

</div>

Financial literacy will be defined as the knowledge and understanding of the matters relating to personal finance. Topics within financial literacy will include, but not be limited to the following: personal budgeting, managing debt, time value of money, investing, tax planning, insurance planning, retirement planning, and estate planning (Lusardi, 2015b).

Financial education will be defined as the process by which financial consumers acquire financial literacy (Lusardi, 2015b).

Financial consumers will be defined as adults who utilize financial products and services as vehicles to plan for and achieve their personal financial goals (Lusardi, 2015b).

Financial goals will be defined as the future conditions that financial consumers desire to satisfy through financial planning (Lusardi, 2015b).

Financial planning will be defined as the process by which financial consumers utilize their acquired financial literacy towards the achievement of their financial goals (Lusardi, 2015b).

Financial services will be defined as the universe of financial products and services that are available to financial consumers as tools to achieve financial goals (Harrison, 2003).

Financial satisfaction will be defined as the self-perception of the ability of financial consumers to achieve and make use of financial literacy in pursuit of their personal financial goals. This self-perception is highly subjective in nature, and thus difficult to measure (Xiao, et al., 2014).

Independent use of the Internet will be defined as people using the Internet to search for the information that interests them and meets their needs (Rubin & Rubin, 2010).

## Summary

As financial literacy and education are important components of modern life, it is important to understand how financial consumers obtain information in this regard. There are a variety of media sources available to obtain investment and financial planning information, and demographic groups may differ in the use of media. Chapter 1 provides an introductory background, the problem statement, and discusses the significance of this research study. The research question that is to be addressed throughout this study is presented along with corresponding hypotheses. Important terms central to this study are defined within the context of the research.

Chapter 2 is a review of the literature that applies to the core concepts of this study, which include financial literacy; information sources and usage; and demographic differences related to gender, age, education, and income level. Chapter 3 is an outline of the research design, methodology of the research process, and protocol for data analysis. Chapter 4 reviews the results of the research process, and Chapter 5 will cover research discussion, conclusions, and implications.

## CHAPTER 2

## REVIEW OF THE LITERATURE

### Introduction

The purpose of this research study is to explore the demographic differences based upon the type of information sourcing used by financial consumers. The scope of the literature review is designed to provide sufficient background to understand the current state of the knowledge set with regards to financial literacy.

The level of financial literacy is not satisfactory in most countries, and the most important means to improve financial literacy is financial education (Fraczek, 2014). Financial education is the process by which financial consumers or investors gain an understanding of financial products, concepts, and risks (Fraczek, 2014). By obtaining information, instruction, and advice, financial consumers develop the skills necessary to be able to make informed choices (Fraczek, 2014). The importance of financial education has increased globally because of the increasing transfer of responsibilities to individuals, and away from traditional entities such as employing corporations and governments (Fraczek, 2014). As such, financial consumers are becoming increasingly responsible for their own financial wellbeing (Fraczek, 2014).

Socio-demographic factors influence the level of financial literacy for financial consumers (Fraczek, 2014). Key demographics for research include gender, age, and life situation demographics such as education level and income level (Fraczek, 2014). Research conclusions may provide answers related to questions involving financial awareness, knowledge, skills, and even attitudes of financial consumers and investors (Fraczek, 2014). Empirical studies indicate that individuals who seek investment and financial planning information are more successful in achieving their financial goals over time (Lusardi & Mitchell, 2007b).

The majority of the literature review includes works that are less than ten years old, and in consideration of the rate of change within the financial services industry, priority was utilized for works that are less than five years old. Peer-reviewed articles that are frequently cited by other researchers were utilized whenever possible. The topics of this literature review include the following: a substantial background on the current state of the knowledge set with regards to financial literacy; an understanding of the available set of investment and financial planning information sources; a review of the current understanding of how information is actually used by the financial consumer; demographic considerations including gender, age, education level, and income level; and a summary that leads into the section on methodology.

## Financial Literacy

The terms financial literacy, financial knowledge, and financial education are used interchangeably in both the literature and popular media and there has been limited effort to differentiate these terms (Huston, 2010). Financial literacy is the set of skills and knowledge necessary to allow an individual to make informed and effective decisions with respect to their financial resources (Lusardi, 2015a). In consideration that many individuals are approaching retirement with low levels of financial wealth, the importance of financial literacy should not be understated (Lusardi, 2015a).

The importance of financial literacy is increasing for financial consumers as a result of the weakening of the government-managed safety net (Xiao, et al., 2014). The Social Security system, which is the traditional retirement income source in the United States, is becoming underfunded (Xiao, et al., 2014). Private-sector pension systems are changing from defined-benefit plans, where the employer is responsible for performance, to defined-contribution plans, where the employee is responsible for performance. As such, financial consumers should worry about their long-term economic security (Xiao, et al., 2014). Unfortunately, empirical research shows that financial consumers are not fully informed and are not able to make proper choices even when appropriate information is available (De Mesa et al., 2008). Lusardi (2015a) recommends that financial literacy instruction be included in high school education curriculums. Of great concern is that organized efforts in support of financial literacy have been surprisingly unsuccessful (Fernandes et al., 2014). While it has been assumed that financial education will result in positive consumer behaviors, it is necessary to design and test financial literacy initiatives for downstream effectiveness (Way, 2014).

Financial literacy can be evaluated on either an objective or a subjective basis. An objective measure would subject financial consumers to a knowledge quiz based upon a specific case study (Xiao, et al., 2014). A subjective measure would ask financial consumers to provide a self-assessment of their financial capability and knowledge (Xiao et al., 2014). In 2009, the Financial Industry Regulatory Authority (FINRA) commissioned the National Financial Capability Study, which was repeated in 2012 (Johnson & Lamdin, 2015). The 2012 National Financial Capability Study had over 25,000 respondents from all fifty states and focused on the demographics of gender, age, education, and income (Johnson & Lamdin, 2015). For this study, financial consumers were subjected to both an objective knowledge-based quiz and subjective self-assessment (Johnson & Lamdin, 2015).

Financial literacy is an even greater issue for complex financial products such as derivative securities, variable annuities, and variable life insurance (Han & Jang, 2013). Some financial products are so complex that even highly educated consumers may not fully understand contract terms thus creating information asymmetry between the provider and the financial consumer (Han & Jang, 2013).

There exists a strong and historical correlation between financial literacy and wealth accumulation as people who better understand their financial environment are more likely to accumulate wealth (Behrman, Mitchell, Soo, & Bravo, 2012). This correlation needs additional analysis as to whether wealth leads to higher levels of financial literacy, or whether financial literacy leads to improved household wealth. The effect of wealth on financial literacy knowledge-based quiz results does show a positive correlation; however, the magnitude of the effect was found to be small (Monticone, 2010). More significant is the financial literacy impact on wealth (van Rooij, Lusardi, & Alessie, 2012). The reason for this is that individuals with good financial literacy are more likely to invest in stocks and are also more likely to plan ahead for retirement (van Rooij et al., 2012).

While it is reassuring that a positive correlation exists between financial literacy and wealth accumulation, the problem of financial security exists for the vast population majority that is not wealthy. This vast population needs access to quality information sources so that they can research, understand, and benefit from their own financial decision making (Lusardi & Mitchell, 2007a). The cost of the lack of financial literacy for individuals and households, that are considered to not be wealthy, is in effect to perpetuate their non-wealthy status over their lifetimes (van Rooij et al., 2012).

**Information Sources**

Financial consumers, through the research of information, may find financial solutions with greater benefits at lower cost, better-managed risk, and greater consumer satisfaction (Lin & Lee, 2004). Consumers tend to research solutions more extensively when purchasing products or services that are expensive and carry risk (Lin & Lee, 2004). As investments can involve substantial amounts of money and risks, information search is an important activity for financial consumers (Lin & Lee, 2004). Consumer confidence relates inversely to the magnitude of the decision, and comfort level drops when the consequences of a poor decision are high and difficult to reverse, thus highlighting the importance of the information search (Loibl, Cho, Diekmann, & Batte, 2009).

Sources of investment and financial planning information include print media, the Internet, radio, and television (Johnson & Lamdin, 2015). Freely-accessible government resources provide financial consumers with high-quality financial information on diverse subjects such as saving, budgeting, college financing, investing, and retirement planning (Rustomfram & Robinson, 2015). Internet-based information sourcing is now globally available, convenient, and cost free (Miller, Kuzner, & Sharma, 2014). A correlation has been established between internet-based information sourcing and consultation with friends and family regarding financial decision-making (Miller et al., 2014).

For young adults, parents and other experienced family members can be an important source of investment and financial planning information (Sherraden & Grinstein-Weiss, 2015). Even social media is now playing a role as an information source (Hean, Worswick, Fenge, Wilkinson, & Fearnley, 2013). The use of social media as an information source is especially prevalent with the Millennial generation (Facebook IQ, 2006).

Professional advisors clearly serve as information sources for investments and financial planning; and financial advice can serve as a substitute for financial literacy (Collins, 2012). What is unknown is the extent to which financial consumers use professional advisors to mitigate their incomplete knowledge (Collins, 2012). Professional advisors engage in the following roles:

offering information, defusing biases that lead to common mistakes, facilitating cognition, overcoming affective issues, and mediating joint decision-making (Collins, 2012). From an economic perspective, a professional advisor may result in lower marginal cost of information sourcing as opposed to searching without assistance (Collins, 2012).

An opportunity for this research study would be to determine which of the various information sources for investment and financial planning information are most popular with financial consumers. This information has potential to be beneficial to financial service providers, government entities, social media providers, and other academic researchers.

**Information Usage**

The benefit of financial literacy is to build the skills and confidence of financial consumers in the use of financial products and services (Fraczek, 2014). This confidence and self-efficacy helps financial consumers to implement financial tasks within their perceptions and abilities (Nilton & Xiao, 2016). A question remains as to whether the information available to financial consumers is making a positive impact regarding financial literacy (Hopkins, Pike, & Littell, 2016). Even when a wrong investment selection has been made, financial consumers are reluctant to use current information to transition towards alternative strategies (Fogel & Berry, 2006).

Information usage has been shown to vary based upon demographics (Ntalianas & Wise, 2011). Women and younger individuals are less likely to utilize educational financial information (Ntalianas & Wise, 2011).

Research establishes a positive correlation between the subjective confidence of financial consumers with respect to their own skills and the probability of seeking professional advice (Robb, Babiarz, & Woodyard, 2012). It is argued that members of the Baby Boomer generation are more knowledgeable, more confident financial consumers, and are more likely to use professional advice in order to avoid the costs associated with poor financial decisions (Robb et al., 2012).

Conversely, the Millennial generation feels disconnected with the professional advice that is provided by the financial services industry (Facebook IQ, 2016). Statistics for the Millennial population show they have an average age of 27, two thirds are college educated, and half earn over $75,000 per year (Facebook IQ, 2016). Millennials tend to be underinvested as they express distrust in financial institutions and the economy in general, yet they recognize that their money could be working harder (Facebook IQ, 2016).

An opportunity for this research study would be to determine if there are differences in sourcing of investment and financial planning information based upon demographics including gender, age, education level, and income level.

## Gender

There are differences between the financial literacy of men and women (Fraczek, 2014). Women are confident in their ability to budget and save money but are less confident than men in their ability to invest (Fraczek, 2014). Women invest less than men in the stock market, and they also score lower on financial literacy (Almenberg & Dreber, 2015).

The 2012 National Financial Capability Study surveyed over 25,000 respondents from all fifty states and segregated the results by gender with 48.6% being male and 51.4% being female (Johnson & Lamdin, 2015). The respondents answered a five-question objective financial literacy quiz, and the results demonstrated that men score higher than women with respect to objective financial literacy (Johnson & Lamdin, 2015). The respondents also answered a two-question subjective financial literacy self-assessment, and the results also demonstrated that men score higher than women with respect to subjective financial literacy (Johnson & Lamdin, 2015).

The financial literacy gender gap between men and women can be partially explained by women's role in household decision-making (Fonseca, Mullen, Zamarro, & Zissimopoulos, 2012). It has been shown that within couples, men are more likely to be identified as the financial representative of the household (Fonseca et al., 2012). However, married and cohabitating individuals do not have higher financial literacy than singles (Fonseca et al., 2012). Of interest is that divorced individuals have lower financial literacy scores than married or never-married respondents (Fonseca et al., 2012). However, divorced individuals tend to improve financial literacy over time as they engage in financial planning for their own benefit (Fonseca et al., 2012). It is important to note that the effects of age and income with regards to financial literacy are not statistically different between men and women (Fonseca et al., 2012). However, higher levels of high school and college education benefit men more than women on financial literacy scoring (Fonseca et al., 2012). Household decision-making roles are thought to partially explain the gender gap, as men tend to make household financial decisions while women tend to focus on other household functions (Fonseca et al., 2012).

The gender gap is especially acute for older women (Lusardi & Mitchell, 2008). Research shows that older women have statistically low levels of financial literacy and that older women tend to not plan adequately for their own retirement (Lusardi & Mitchell, 2008). A reason for this may be that higher rates of divorce and lower remarriage are increasing the percentage of women that are approaching retirement unmarried. If the husband served as the household financial decision maker, the wife may have insufficient financial literacy to plan for her individual retirement (Fonseca et al., 2012).

An opportunity for this research study would be to determine if there are differences in sourcing of investment and financial planning information based upon gender. If so, there may be means identified to help close the financial literacy gap between men and women.

## Age

The level of financial literacy for financial consumers correlates with age (Fraczek, 2014). This correlation is positive as individual consumers' age through the life cycle, but reverses near the end of the life cycle (Fraczek, 2014). Older financial consumers demonstrate increased levels of both objective and subjective financial literacy, along with perceived financial capability, as compared to younger financial consumers (Xiao, Chen, & Sun, 2015). However, this benefit tends to reverse late in the life cycle possibly as a result of decreasing mental capability (Lusardi, 2012).

The 2012 National Financial Capability Study surveyed over 25,000 respondents from all fifty states and segregated the results by age categories (Johnson & Lamdin, 2015). The age categories and percentage of respondents were the following: age 18 to 34 (30.6%), age 35-54 (35.9%), and 33.4% of the respondents reported themselves as age 55 and over (Johnson & Lamdin, 2015). The respondents answered a five-question objective financial literacy quiz, and the results confirmed a positive correlation between age and objective financial literacy (Johnson & Lamdin, 2015). The respondents also answered a two-question subjective financial literacy self-assessment, and the results also confirmed a positive correlation between age and subjective financial literacy (Johnson & Lamdin, 2015). It should be noted that the 2012 National Financial Capability Study did not segregate respondents that were age 80 and over, so the effect of declining financial literacy that occurs near the end of the life cycle was not observed in the study (Fraczek, 2014).

The reason for the positive correlation between age and financial literacy is due to the natural development of financial self-efficacy for individuals that occurs over time (Xiao, et al., 2015). Self-efficacy is an important factor in the psychological influence of human behaviors (Xiao, et al., 2015). ). As financial consumers age, their financial self-efficacy should increase along with their more complicated financial lives (Xiao et al., 2015). This process continues until about age 80 when decreasing mental capability offsets increasing self-efficacy (Xiao, et al., 2015).

An opportunity for this research study would be to determine if there are differences in sourcing of investment and financial planning information based upon age. If so there may be means identified to help improve financial literacy at younger age levels. For example, the Millennial generation has been referred to as digital natives (Prensky, 2001). This generation, born digital, can be expected to learn differently than prior generations who learned to use computers later in adulthood (Prensky, 2001).

## Education

The level of financial literacy for financial consumers is positively correlated with educational attainment Fraczek, 2014). The 2012 National Financial Capability Study surveyed over 25,000 respondents from all fifty states and segregated the results by education level (Johnson & Lamdin, 2015). The education level categories and percentage of respondents were the following: did not complete high school (8.7%), high school graduate (29.4%), some college (35.9%), college graduate (16.1%), and 9.9% of the respondents reported postgraduate education (Johnson & Lamdin, 2015). The respondents answered a five-question objective financial literacy quiz, and the results confirmed a positive correlation between education level and objective financial literacy (Johnson & Lamdin, 2015). The respondents also answered a two-question subjective financial literacy self-assessment, and the results also confirmed a positive correlation between education level and subjective financial literacy (Johnson & Lamdin, 2015).

The reason for the positive correlation between education level and financial literacy may have to do with other factors such as income, as college graduates earn more than individuals without a college degree (Monticone, 2010). However, financial literacy may depend upon cognitive ability and other controls such as the willingness to acquire knowledge (Monticone, 2010). While a positive correlation does exist between education level and financial literacy, the theory behind this correlation needs to be better understood (Monticone, 2010).

An opportunity for this research study would be to determine if there are differences in sourcing of investment and financial planning information based upon education level. If so, there may be means identified to improve financial literacy for those who do not have the benefit of a college education.

## Income

The level of financial literacy for financial consumers is positively correlated with income level (Fraczek, 2014). The 2012 National Financial Capability Study surveyed over 25,000 respondents from all fifty states and segregated the results by income level (Johnson & Lamdin, 2015). The income level categories and percentage of respondents were the following: under $25,000 (26.5%), $25,000 to $50,000 (26.1%), $50,000 to $100,000 (30.3%), and 16.9% of the respondents reported over $100,000 in annual income (Johnson & Lamdin, 2015). The respondents answered a five-question objective financial literacy quiz, and the results confirmed a positive correlation between income level and objective financial literacy (Johnson & Lamdin, 2015). The respondents also answered a two-question subjective financial literacy self-assessment, and the results also confirmed a positive correlation between income level and subjective financial literacy (Johnson & Lamdin, 2015).

The reason for the positive correlation between income level and financial literacy is two-fold. First lower income individuals may not have surplus financial resources that can be set away to meet future financial planning needs, as many households are struggling financially (Sherraden & Grinstein-Weiss, 2015). It is an unfortunate reality that individuals who lack control of their financial environment are unlikely to build financial literacy (Behrman et al., 2012). The other factor relates to financial self-efficacy in that as the income of individuals increases, financial literacy would be expected to increase alongside their more complicated financial lives (Xiao, et al., 2015). Experiential learning may be the reason for the positive correlation between income level and financial literacy (Johnson & Lamdin, 2015).

An opportunity for this research study would be to determine if there are differences in sourcing of investment and financial planning information based upon income level. If so there may be means identified to improve financial literacy for low to moderate income levels.

## Summary

As reviewed in this chapter, the basis of this research topic is to understand the state of the current knowledge set with regards to financial literacy. While it had been thought that financial literacy could be taught, recent research indicates that experiential learning is more effective in the development of financial literacy (Johnson & Lamdin, 2015). Experiential learning involves active information search by financial consumers (Johnson & Lamdin, 2015). Various sources for investment and financial planning information are reviewed. These information sources can be used as tools to help financial consumers modify their behaviors with regards to investing and financial planning (Lusardi & Mitchell, 2007b). Differences in financial literacy exist within demographics based upon gender, age, education, and income (Johnson & Lamdin, 2015). This research project will attempt to explore the demographic differences based upon the mode of information sourcing used by the financial consumer.

Higher levels of financial literacy allow financial consumers to aspire to change their financial behavior (Fraczek, 2014). There is a positive association between financial literacy and financial satisfaction (Xiao, et al., 2014). There is also a positive association between financial satisfaction and quality of life (Michalos, 2008). Unfortunately, there is no consensus on the best way to measure financial satisfaction (Joo & Grable, 2004). However, the aim of this research study will be to provide insight on how financial consumers seek the information they need to improve their financial literacy. This insight will be potentially valuable to financial service providers, government entities, social media providers, academic researchers, and to the investing public in general.

Chapter 3 will provide a summary of research methodology to be applied by this research study. Sample selection, data collection methods, and data management procedures will be discussed. A review of the statistical and data analytics used to test the hypotheses of the research study will also be included in the next chapter.

# CHAPTER 3

# METHODOLOGY

## Introduction

The literature review has identified demographic differences regarding financial literacy arising from differences in gender, age, education-level, and income-level among financial consumers. The purpose of this research project is to gain insight and understanding regarding these differences.

This chapter reviews the research question, provides the research design, describes the population and sample, presents limitations and assumptions, and reviews how data collection and analysis were accomplished. Validity, reliability, and ethical considerations are also addressed. The research question is as follows:

**RQ1.   How are gender, age, education-level, and income-level associated with independent use of the Internet, versus the use of professional advisors, for obtaining investment and financial planning information?**

## Research Design

The research is designed to study the four demographic groups. The demographic group of gender is defined as female, male, and those that may identify as other. To the extent that there may be transgender participants, the study will accept the gender that they associate with. The demographic group of age is segregated into four ranges: 18 to 34, 35 to 54, 55 to 74, and 75 up. The age 18 lower bound represents adulthood when individuals can independently hold financial accounts.  These age ranges span the Millennials, Generation X, the Baby Boomers, and their parents. The demographic group of education level is simply defined as college educated, or not. The demographic group for income level is segregated into three ranges: under $25,000 per year, $25,000 to $100,000 per year, and above $100,000 per year. The $25,000 per year threshold represents the 2017 Federal poverty level for a family of four (U.S. Department of Health and Human Services, 2017). The figure of $100,000 was chosen as a threshold based upon interviews with financial services corporate executives who identified that level as the target market for their firm's wealth management services (J. Caven, personal interview, April 17, 2017; J. Kramer, personal interview, August 14, 2017). The four demographic groups are the independent variables.

The research is designed to evaluate two primary types of information sources as means of obtaining investment and financial planning information. The two primary types of information sources are the Internet and professional advisors. Professional advisors are defined as attorneys, accountants, investment advisors, and financial planners. The two types of information sources are the dependent variables.

The research is intended to provide quantitative discussion to the following sixteen hypotheses:

**H1a.    There is a statistically significant difference in use of the Internet to gather investment and financial planning information based upon gender.**

**H1b.    There is a statistically significant difference in frequency of Internet use to gather investment and financial planning information based upon gender.**

**H1c.    There is a statistically significant difference in use of professionals to gather investment and financial planning information based upon gender.**

**H1d.    There is a statistically significant difference in frequency of professional use to gather investment and financial planning information based upon gender.**

**H2a.** There is a statistically significant difference in use of the Internet to gather investment and financial planning information based upon age.

**H2b.** There is a statistically significant difference in frequency of Internet use to gather investment and financial planning information based upon age.

**H2c.** There is a statistically significant difference in use of professionals to gather investment and financial planning information based upon age.

**H2d.** There is a statistically significant difference in frequency of professional use to gather investment and financial planning information based upon age.

**H3a.** There is a statistically significant difference in use of the Internet to gather investment and financial planning information based upon education level.

**H3b.** There is a statistically significant difference in frequency of Internet use to gather investment and financial planning information based upon education level.

**H3c.** There is a statistically significant difference in use of professionals to gather investment and financial planning information based upon education level.

**H3d.** There is a statistically significant difference in frequency of professional use to gather investment and financial planning information based upon education level.

**H4a.** There is a statistically significant difference in use of the Internet to gather investment and financial planning information based upon income level.

**H4b.** There is a statistically significant difference in frequency of Internet use to gather investment and financial planning information based upon income level.

**H4c.** There is a statistically significant difference in use of professionals to gather investment and financial planning information based upon income level.

**H4d.    There is a statistically significant difference in frequency of professional use to gather investment and financial planning information based upon income level.**

## Population and Sample

The survey population is defined as the author's social media connections on LinkedIn (N>2300). This population primarily consists of educated professionals, who are either working or retired, and who can be expected to be consumers of investment information as they plan for strategic goals for themselves and their families. This population is expected to provide adequate demographic distribution for gender, age, education-level, and income-level. There will be no need to select a sample due to the manageable size of the population. A response rate of over 5 percent should provide adequate data for the study.

## Limitations and Assumptions

The research study is dependent upon the propensity of the survey population to participate in the survey. The ease of the survey questionnaire should help facilitate an adequate response rate. The challenge is to create an interest level for the survey that exceeds the 'background noise level' that the target population is subjected to on a daily basis.

This study has several limitations. As the data is collected through the use of a voluntary survey questionnaire, the participant response is biased towards those in the population who are inclined to answer surveys. Further opportunity for bias is introduced through the selection of the survey population. This population is defined as the author's social media connections on LinkedIn (N > 2300). As such, the respondents know the identity and affiliations of the author. While this familiarity helps to facilitate an excellent response rate, the familiarity also creates opportunity for bias in the responses.

In addition, it should be expected that the survey population does not geographically represent the national population. As the survey was directed towards working and retired professionals that maintain profiles on LinkedIn, it should further be expected that the survey population does not reflect the national population on either an economic or an educational basis. There is no expectation that the survey reflects the population of LinkedIn users as a whole. As such, the survey population may be too uniform to provide statistically significant outcomes for the research question and hypotheses. It should be expected that the survey population would not be generalizable to a larger population.

### Data Collection

The survey questionnaire can be found in Appendix A. An invitation to participate was distributed with a link to the survey. This communication identified the survey with the Doctor of Business Administration program at Wilmington University, addressed the purpose of the study, provided a time estimate to complete the survey, addressed anonymity, and requested participation. The survey was administered using a paid subscription to Survey Monkey. The paid subscription allowed for the full survey to be administered including demographic information and for the data to be downloaded for analysis.

The survey population was provided with an invitation to participate with an internet link that directed them to the survey questionnaire. The invitation to participate had 1302 views on LinkedIn. Of these views, 246 people clicked through to the survey and 233 people answered at least one survey question. The survey was restricted to allow only one response per internet protocol address. The survey period was closed after seven days, at which point the data was downloaded, scrubbed, coded, and analyzed.

## Data Analysis

The study produced a total of 233 responses. The policy to scrub responses where the participant failed to answer all 9 closed-end questions yielded a usable data set of 214 responses. The responses were then coded on a five-point scale for the Likert questions, and on a two, three, or four-point scale as appropriate for the four demographic questions. The open-end question was categorized to look for patterns as described in the findings.

The demographics included the following four groupings: gender (female, male, or other), education (college graduate or not), age (18 to 34, 35 to 54, 55 to 74, and 75 up), and household income (below $25,000, $25,000 to $100,000, and above $100,000). Descriptive statistics are provided for the four groupings. Demographic analysis was not run unless there were at least 10 respondents for a grouping option. For example, there would need to be 10 male and 10 female respondents to run gender analysis. Demographic analysis will determine if there are different answers to the research questions based upon the groupings selected for this study. Group analysis will provide insight into areas for future research.

Data was analyzed using IBM SPSS software. The independent variables were reviewed for bias, and the dependent variables were reviewed for normality. As the dependent variables are ordinal, and therefore non-normal, non-parametric analysis was performed for the four dependent variables. The goal of the data analysis is to determine if the hypotheses are either supported or not supported by the study.

## Validity and Reliability

The success of this research study depends upon both validity and reliability of the data that is collected. Reliability is the extent to which a survey instrument produces stable and consistent results (Pearson, 2010). As this study was not time-based, there was no need to measure test-retest reliability. As there was a single data evaluator, there was no need to measure interrater reliability, and as there was a single survey instrument, there was no need to measure parallel form reliability.

Internal validity is the extent to which the data measures what it claims to measure (Pearson, 2010). This research study is considered internally valid based upon dissertation committee consensus that face validity exists for the survey questions. External validity is the extent to which the results can be generalized beyond the immediate study (Pearson, 2010). This research study cannot be expected to be perfectly generalized to the total population of financial consumers. While the results may be representative, a more encompassing study would be needed to better address external validity.

## Ethical Considerations

The Wilmington University Human Subjects Review Committee has recorded and reviewed this research protocol. Participation in the survey is both voluntary and anonymous. The data collected from individual survey responses cannot put participants at risk in any way. There were not any benefits or compensation offered to the participants, and there are no known conflicts of interest between the participants and the administrators of the survey instrument.

The survey request for demographic information is sufficiently broad to protect the respondents' anonymity. All published data is in aggregate form, which precludes the possibility of disclosure of individual proprietary information.

## Summary

An overview of the research design and methodology used for this research study has been provided within this chapter. The purpose of this research study is to understand the differences between the demographic categories of gender, age, education-level, and income-level; with the means of obtaining investment and financial planning information. In order to gain insight into these relationships, a primary data analysis was performed. The data set was comprised of 233 respondents to a survey that was made available on social media, and the data was analyzed using IBM SPSS software. The results of this research study are provided in Chapter 4.

# CHAPTER 4

# RESULTS

## Introduction

Chapter 4 presents empirical results based upon the methodology prescribed in Chapter 3. The order of this chapter describes the respondent population, states the research question and hypothesis, presents the statistical results, and provides a summary of where the null hypothesis is either rejected or fails to be rejected.

## Description of the Population

There were 233 respondents to the survey instrument. The policy to only include respondents who answered all nine closed-end questions yielded a data set of 214 responses. Population demographics were evaluated based upon gender, age, education, and household income.

Population frequency by gender shows a bias towards male respondents (see Table 1). As demographic analysis will only be run where there are at least 10 respondents for a grouping option, the gender group of "other" was not separately evaluated in this study.

Table 1

*Frequency Distribution by Gender*

| Gender | Frequency | Percent |
|--------|-----------|---------|
| Male   | 148       | 69.2    |
| Female | 63        | 29.4    |
| Other  | 3         | 1.4     |
| Total  | 214       | 100.0   |

Population frequency by age range shows a concentration for the 35-54 and 55-74 ranges (see Table 2). There were sufficient responses for all four groups to be evaluated.

Table 2

*Frequency Distribution by Age*

| Age Range | Frequency | Percent |
| --- | --- | --- |
| 18-34 | 18 | 8.4 |
| 35-54 | 76 | 35.5 |
| 55-74 | 104 | 48.6 |
| 75+ | 16 | 7.5 |
| Total | 214 | 100.0 |

Population frequency by education level shows a bias towards college graduates (see Table 3). There were sufficient responses for both groups to be evaluated.

Table 3

*Frequency Distribution by Education*

| College Graduate | Frequency | Percent |
| --- | --- | --- |
| Yes | 186 | 86.9 |
| No | 28 | 13.1 |
| Total | 214 | 100.0 |

Population frequency by household income shows a bias towards households with income exceeding $100,000 per year (see Table 4). There were sufficient responses for all three groups to be evaluated.

Table 4

*Frequency Distribution by Household Income*

| Annual Income | Frequency | Percent |
| --- | --- | --- |
| Under $25,000 | 19 | 8.9 |
| $25,000 to $100,000 | 73 | 34.1 |
| Over $100,000 | 122 | 57.0 |
| Total | 214 | 100.0 |

## Research Question and Hypothesis

This study seeks to answer the following research question that evaluates how financial consumers obtain investment and financial planning information based upon the four key demographics of gender, age, education-level, and income-level:

**RQ1.   How are gender, age, education-level, and income-level associated with independent use of the Internet, versus the use of professional advisors, for obtaining investment and financial planning information?**

The research is intended to provide quantitative discussion to the following sixteen hypotheses:

**H1a.     There is a statistically significant difference in use of the Internet to gather investment and financial planning information based upon gender.**

**H1b.     There is a statistically significant difference in frequency of Internet use to gather investment and financial planning information based upon gender.**

**H1c.** There is a statistically significant difference in use of professionals to gather investment and financial planning information based upon gender.

**H1d.** There is a statistically significant difference in frequency of professional use to gather investment and financial planning information based upon gender.

**H2a.** There is a statistically significant difference in use of the Internet to gather investment and financial planning information based upon age.

**H2b.** There is a statistically significant difference in frequency of Internet use to gather investment and financial planning information based upon age.

**H2c.** There is a statistically significant difference in use of professionals to gather investment and financial planning information based upon age.

**H2d.** There is a statistically significant difference in frequency of professional use to gather investment and financial planning information based upon age.

**H3a.** There is a statistically significant difference in use of the Internet to gather investment and financial planning information based upon education level.

**H3b.** There is a statistically significant difference in frequency of Internet use to gather investment and financial planning information based upon education level.

**H3c.** There is a statistically significant difference in use of professionals to gather investment and financial planning information based upon education level.

**H3d.** There is a statistically significant difference in frequency of professional use to gather investment and financial planning information based upon education level.

**H4a.**   **There is a statistically significant difference in use of the Internet to gather investment and financial planning information based upon income level.**

**H4b.**   **There is a statistically significant difference in frequency of Internet use to gather investment and financial planning information based upon income level.**

**H4c.**   **There is a statistically significant difference in use of professionals to gather investment and financial planning information based upon income level.**

**H4d.**   **There is a statistically significant difference in frequency of professional use to gather investment and financial planning information based upon income level.**

## Results

Four dependent variables were evaluated for each of the four demographic categories. The dependent variables were the participant's responses to the following four closed-end questions.

Q1.  I find the Internet to be very useful for obtaining investment and financial planning information. (Strongly agree / Agree / Neither agree or disagree / Disagree / Strongly disagree)

Q2.  How frequently do you use the Internet to search for and obtain investment and financial planning information? (Daily / Weekly / Monthly / Quarterly / Annually / Never)

Q3.  I find consulting professionals (attorneys, CPAs, investment advisors, financial planners) to be very useful for obtaining investment and financial planning information. (Strongly agree / Agree / Neither agree or disagree / Disagree / Strongly disagree)

Q4.  How frequently do you consult professionals to obtain investment and financial planning information? (Daily / Weekly / Monthly / Quarterly / Annually / Never)

These four dependent variables are referred to as Internet use, Internet frequency, professional use, and professional frequency. Reverse coding was used so that "strongly agree" and "daily" were scored as 5 on the Likert scale. Visual analysis of the dependent variable distributions showed obvious skewness and kurtosis. Additionally, as the dependent variables are ordinal, non-parametric analysis was used. The Wilcoxon signed-rank test was used as the non-parametric alternative to the Student's t-test, and the Welch test was used as the non-parametric alternative to ANOVA.

The four demographic categories were analyzed in the stated order of the hypotheses. A Wilcoxon signed-rank test was run for the two groups, male and female, within the category of gender (see table 5). Men are more likely to use the Internet ($p = .002$) and more frequently use the Internet ($p = .000$) than women for obtaining investment and financial planning information. This finding is consistent with Fonseca et al. (2012) in that men are more likely to be identified as the financial representative of the household. However, women are more likely to consult professionals ($p = .039$) than men. This finding is consistent with Fraczek (2014). The magnitude of the mean difference was most pronounced for Internet frequency (.827) whereas men tend to search the Internet monthly and women tend to search the Internet on a quarterly basis. The magnitude of the mean difference for Internet use and professional use are less meaningful, although statistically significant. Within the four dependent variables, the null hypothesis is rejected for Internet use, Internet frequency, and professional use, but fails to be rejected for professional frequency. For the category of gender, the null hypothesis is rejected for H1a, H1b, and H1c, but fails to be rejected for H1d.

Table 5

*Wilcoxon Test for Gender*

| | Wilcoxon Significance | Mean Difference | Std. Error Difference |
|---|---|---|---|
| Internet Use | .002 | .460 | .148 |
| Internet Frequency | .000 | .827 | .226 |
| Professional Use | .039 | -.319 | .153 |
| Professional Frequency | .418 | .131 | .161 |

A Welch test was run for the four groups within the demographic category of age (see table 6). There were significant differences with three of the four dependent variables; Internet use ($p$ = .006), Internet frequency ($p$ = .002), and professional frequency (p = .000). Games-Howell post-hoc testing was used to show differences between age groups (see table 7). The age group of 75+ was less likely to use the Internet for obtaining investment and financial planning information than the other three age groups. The 55-74 and 75+ age groups consulted professionals more frequently than the 18-34 and 35-54 age groups. It should be noted that the magnitude of the mean difference between the 55-74 and 35-54 for professional frequency was small (.470), so although this difference was statistically significant, the difference is not necessarily meaningful. These findings are consistent with Fraczek (2014) in that financial literacy has a positive correlation with age through the life cycle but reverses near the end of the life cycle possibly because of declining mental capability. Within the four dependent variables, the null hypothesis is rejected for Internet use, Internet frequency, and professional frequency; but fails to be rejected for professional use.  For the category of age, the null hypothesis is rejected for H2a, H2b, and H2d, but fails to be rejected for H2c.

Table 6

*Welch Test for Age Ranges*

|  | Welch Statistic | Significance |
| --- | --- | --- |
| Internet use | 4.785 | .006 |
| Internet frequency | 5.611 | .002 |
| Professional use | 2.411 | .080 |
| Professional frequency | 8.695 | .000 |

Table 7

*Games-Howell post-hoc Testing for Age Groups*

|  | Age Group | Comparison Group | Mean Difference | Significance |
|---|---|---|---|---|
| Internet use | 18-34 | 75+ | 1.063 | .045 |
|  | 35-54 | 75+ | .997 | .012 |
|  | 55-74 | 75+ | 1.072 | .006 |
| Internet frequency | 35-54 | 75+ | 1.016 | .035 |
|  | 55-74 | 75+ | 1.389 | .002 |
| Professional frequency | 55-74 | 18-34 | 1.043 | .003 |
|  | 55-74 | 35-54 | .470 | .024 |
|  | 75+ | 18-34 | 1.514 | .001 |
|  | 75+ | 35-54 | .941 | .015 |

A Wilcoxon signed-rank test was run for the two groups within the category of education-level (see table 8). The data did not reveal significant differences between Internet use, Internet frequency, and professional use versus education-level. However, the data reveals a significant difference between education-level and professional frequency ($p = .007$). College graduates more frequently consult professionals for obtaining investment and financial planning information than those respondents that did not have a college degree. This finding is consistent with Johnson and Lamdin (2015) who identified positive correlation between education-level and financial literacy. The magnitude of the mean difference for professional frequency (.664) shows that college graduates consult professionals quarterly to annually while the respondents that did not have a college degree consult professionals annually or less. Within the four dependent variables, the null hypothesis is rejected for professional frequency, but fails to be rejected for Internet use, Internet frequency, and professional use.   For the category of education-level, the null hypothesis is rejected for H3d, but fails to be rejected for H3a, H3b and H3c.

Table 8

*Wilcoxon Test for Education Level*

|  | Wilcoxon Significance | Mean Difference | Std. Error Difference |
|---|---|---|---|
| Internet use | .061 | .503 | .259 |
| Internet frequency | .456 | .264 | .350 |
| Professional use | .285 | .272 | .251 |
| Professional frequency | .007 | .664 | .231 |

A Welch test was run for the three groups within the demographic category of income-level (see table 9). The data did not reveal significant differences between Internet use, professional use, and professional frequency versus income-level. However, the data reveals a significant difference between income-level and Internet frequency ($p = .050$). This finding is consistent with prior work by Fraczek (2014) in which there was positive correlation with income-level. Games-Howell post-hoc testing was run for Internet frequency but failed to show any statistical differences between income-level groups (see table 10). Within the four dependent variables, the null hypothesis is rejected for Internet frequency but fails to be rejected for Internet use, professional use, and professional frequency. For the category of income-level, the null hypothesis is rejected for H4b, but fails to be rejected for H4a, H4c, and H4d.

Table 9

*Welch Test for Income Level*

|  | Welch Statistic | Significance |
|---|---|---|
| Internet use | .520 | .598 |
| Internet frequency | 3.170 | .050 |
| Professional use | 1.003 | .374 |
| Professional frequency | 2.629 | .082 |

Table 10

*Games-Howell post-hoc Testing for Income Level*

|  | Income Group | Comparison Group | Mean Difference | Significance |
|---|---|---|---|---|
| Internet frequency | Under $25,000 | $25,000 to $100,000 | -.498 | .395 |
|  | Under $25,000 | Over $100,000 | -.850 | .066 |
|  | $25,000 to $100,000 | Over $100,000 | -.352 | .294 |

**Summary**

The purpose of this project was to determine if statistical differences exist between the demographic categories that were researched in this study regarding the respondent's independent use of the Internet, versus the use of professional advisors, for obtaining investment and financial planning information. The four demographic categories studied were gender, age, education-level, and income-level.

For the demographic category of gender, there were significant differences with three of the dependent variables. Men are more likely to use the Internet and more frequently use the Internet than women for obtaining investment and financial planning information. However, women are more likely to consult professionals than men. The null hypothesis is rejected for H1a, H1b, and H1c but fails to be rejected for H1d.

For the demographic category of age, there were significant differences with three of the four dependent variables. The age group of 75+ was less likely to use the Internet for obtaining investment and financial planning information than the other three age groups. The 55-74 and 75+ age groups consulted professionals more frequently than the 18-34 and 35-54 age groups. The null hypothesis is rejected for H2a, H2b, and H2d but fails to be rejected for H2c.

For the education-level demographic category, there was a significant difference with the dependent variable for frequency of consulting professionals. College graduates more frequently consult professionals for obtaining investment and financial planning information than those respondents that did not have a college degree. The null hypothesis is rejected for H3d but fails to be rejected for H3a, H3b and H3c.

For the demographic category of income-level, there was a significant difference with Internet frequency. Respondents with higher incomes tend to use the Internet more frequently for obtaining investment and financial planning information. The null hypothesis is rejected for H4b but fails to be rejected for H4a, H4c, and H4d.

      Chapter 5 will provide discussion, conclusions, and implications of these results.

# CHAPTER 5

# DISCUSSION, CONCLUSIONS, AND IMPLICATIONS

## Overview

In Chapter 5, there is discussion of key elements from the literature review, a synopsis of the sample population, an analysis of the data, and a summary of conclusions for the research question and associated hypotheses. Implications for commercial practice are identified along with specific recommendations. The limitations of this research project are stated along with recommendations for future research.

## Key Elements from the Literature Review

Financial literacy and financial education are important components of modern life. Nearly every adult person in the United States, and also the world, needs at least a basic level of financial literacy to be able to function in society (Johnson & Lamdin, 2015).

The level of financial literacy is not satisfactory in most countries, and the most important means to improve financial literacy is financial education (Fraczek, 2014). Financial education is the process by which financial consumers or investors gain an understanding of financial products, concepts, and risks (Lusardi, 2015a. By obtaining information, instruction, and advice, financial consumers develop the skills necessary to be able to make informed choices (Fraczek, 2014). The importance of financial education has increased globally because of the increasing transfer of responsibilities to individuals and away from traditional entities such as employing corporations and governments (Fraczek, 2014). As such, financial consumers are becoming increasingly responsible for their own financial wellbeing (Johnson & Lamdin, 2015).

The purpose of this research study is to explore the demographic differences based upon the type of information sourcing used by financial consumers. The scope of the literature review is designed to provide sufficient background to understand the current state of the knowledge set with regards to financial literacy.

It should be recognized that socio-demographic factors have influence over the level of financial literacy for financial consumers (Fraczek, 2014). Key demographics for research include gender, age, and life situation demographics such as education level and income level (Fraczek, 2014). Research conclusions may provide answers related to questions involving financial awareness, knowledge, skills, and even attitudes of financial consumers and investors (Fraczek, 2014). Empirical studies indicate that individuals who seek investment and financial planning information are more successful in achieving their financial goals over time (Lusardi & Mitchell, 2007b).

With respect to gender, there are differences between the financial literacy of men and women (Fraczek, 2014). Women are confident in their ability to budget and save money but are less confident than men in their ability to invest (Fraczek, 2014). Women invest less than men in the stock market, and they also score lower on financial literacy (Almenberg & Dreber, 2015). The financial literacy gender gap between men and women can be partially explained by women's role in household decision-making (Fonseca et al., 2012). It has been shown that within couples, men are more likely to be identified as the financial representative of the household (Fonseca et al., 2012).

With respect to age, the level of financial literacy for financial consumers correlates with age (Fraczek, 2014). This correlation is positive as individual consumers' age through the life cycle, but reverses near the end of the life cycle (Fraczek, 2014). Older financial consumers demonstrate increased levels of both objective and subjective financial literacy, along with perceived financial capability, as compared to younger financial consumers (Xiao, et al., 2015).

However, this benefit tends to reverse late in the life cycle possibly as a result of decreasing mental capability (Lusardi, 2012). The reason for the positive correlation between age and financial literacy is due to the natural development of financial self-efficacy for individuals that occurs over time (Xiao, et al., 2015). Self-efficacy is an important factor in the psychological influence of human behaviors (Xiao, et al., 2015). As financial consumers age, their financial self-efficacy should increase along with their more complicated financial lives (Xiao, et al., 2015). This process continues until about age 80 when decreasing mental capability offsets increasing self-efficacy (Xiao, et al., 2015).

The level of financial literacy for financial consumers is positively correlated with educational attainment (Fraczek, 2014). The reason for the positive correlation between education level and financial literacy may have to do with other factors such as income, as college graduates earn more than individuals do without a college degree (Monticone, 2010). However, financial literacy may depend upon cognitive ability and other controls such as the willingness to acquire knowledge (Monticone, 2010). While a positive correlation does exist between education level and financial literacy, the theory behind this correlation needs to be better understood (Monticone, 2010).

The level of financial literacy for financial consumers is positively correlated with income level (Fraczek, 2014). The reason for the positive correlation between income level and financial literacy is two-fold. First lower income individuals may not have surplus financial resources that can be set away to meet future financial planning needs, as many households are struggling financially (Sherraden & Grinstein-Weiss, 2015). It is an unfortunate reality that individuals who lack control of their financial environment are unlikely to build financial literacy (Behrman et al., 2012). The other factor relates to financial self-efficacy in that as the income of individuals increases, financial literacy would be expected to increase alongside their more complicated financial lives (Xiao, et al., 2015). Experiential learning may be the reason for the positive correlation between income level and financial literacy (Johnson & Lamdin, 2015).

As reviewed in Chapter 2, the basis of this research topic is to understand the state of the current knowledge set with regards to financial literacy. While it had been thought that financial literacy could be taught, recent research indicates that experiential learning is more effective in the development of financial literacy (Johnson & Lamdin, 2015). Differences in financial literacy exist within demographics based upon gender, age, education, and income (Johnson & Lamdin, 2015). This research project attempted to explore the demographic differences based upon the mode of information sourcing used by the financial consumer.

### Sample Population

The survey population was defined as the author's social media connections on LinkedIn (N>2300). This population primarily consists of educated professionals, who are either working or retired, and who can be expected to be consumers of investment information as they plan for strategic goals for themselves and their families. This population was expected to provide adequate demographic distribution for gender, age, education-level, and income-level.

The survey population was provided with an invitation to participate with an internet link that directed them to the survey questionnaire. The invitation to participate had 1302 views on LinkedIn. Of these views, 246 people clicked through to the survey and 233 people answered at least one survey question. The survey was restricted to allow only one response per internet protocol address.  The survey period was closed after seven days.

## Analysis of the Data

The study produced a total of 233 responses. The policy to scrub responses where the participant failed to answer all nine closed-end questions yielded a usable data set of 214 responses. The demographics included the following four groupings: gender (female, male, or other), education (college graduate or not), age (18 to 34, 35 to 54, 55 to 74, and 75 up), and household income (below $25,000, $25,000 to $100,000, and above $100,000). Descriptive statistics were provided for the four groupings. Demographic analysis was not run unless there were at least 10 respondents for a grouping option. For example, there would need to be 10 male and 10 female respondents to run gender analysis. The only demographic category that had fewer than 10 respondents was the gender category of "other". Demographic analysis determined if there were different answers to the research questions based upon the groupings selected for this study. Data was analyzed using IBM SPSS software. The independent variables were reviewed for bias, and the dependent variables were reviewed for normality. As the dependent variables were ordinal, non-parametric analysis was performed for the four dependent variables. The goal of the data analysis was to determine if the hypotheses were either supported or not supported by the study.

## Conclusions

The purpose of this project was to determine if statistical differences exist between the demographic categories that were researched in this study regarding the respondent's independent use of the Internet, versus the use of professional advisors, for obtaining investment and financial planning information. The four demographic categories studied were gender, age, education-level, and income-level.

For the demographic category of gender, there were significant differences with three of the dependent variables. Men are more likely to use the Internet and more frequently use the Internet than women for obtaining investment and financial planning information. However, women are more likely to consult professionals than men. Within the four dependent variables, the null hypothesis is rejected for Internet use, Internet frequency, and professional use. This finding is consistent with Fonseca et al. (2012) in that men are more likely to be identified as the financial representative of the household. Women are confident in their ability to budget and save money but are less confident than men in their ability to invest (Fraczek, 2014). Women invest less than men in the stock market, and they also score lower on financial literacy (Almenberg & Dreber, 2015). The null hypothesis fails to be rejected for professional frequency, as a statistical difference was not found for this variable.

For the demographic category of age, there were significant differences with three of the four dependent variables. The age group of 75+ was less likely to use the Internet for obtaining investment and financial planning information than the other three age groups. The 55-74 and 75+ age groups consulted professionals more frequently than the 18-34 and 35-54 age groups. Within the four dependent variables, the null hypothesis is rejected for Internet use, Internet frequency, and professional frequency. These findings are consistent with Fraczek (2014) in that financial literacy has a positive correlation with age through the life cycle but reverses near the end of the life cycle possibly because of declining mental capability. The reason for the positive correlation between age and financial literacy is due to the natural development of financial self-efficacy for individuals that occurs over time (Xiao, et al., 2015). Self-efficacy is an important factor in the psychological influence of human behaviors (Xiao, et al., 2015). As financial consumers age, their financial self-efficacy should increase along with their more complicated financial lives. The null hypothesis fails to be rejected for professional use, as a statistical difference was not found for this variable.

For the education-level demographic category, there was a significant difference with the dependent variable of the frequency of consulting professionals. College graduates more frequently consult professionals for obtaining investment and financial planning information than those respondents that did not have a college degree. The null hypothesis is rejected for this variable. This finding is consistent with Johnson and Lamdin (2015) who identified positive correlation between education-level and financial literacy. The reason for this is that financial literacy may depend upon cognitive ability and other controls such as the willingness to acquire knowledge (Monticone, 2010). The null hypothesis fails to be rejected for Internet use, Internet frequency, and professional frequency, as statistical differences were not found for these variables.

For the demographic category of income-level, there was a significant difference with Internet frequency. Respondents with higher incomes tend to use the Internet more frequently for obtaining investment and financial planning information. The null hypothesis is rejected for this variable. This finding is consistent with prior work by Fraczek (2014) in which there was positive correlation with income-level. The reason for the expected positive correlation between income-level and financial literacy is two-fold. First lower income individuals may not have surplus financial resources that can be set away to meet future financial planning needs. The other factor relates to financial self-efficacy in that as the income of individuals increases, financial literacy would be expected to increase alongside their more complicated financial lives. The null hypothesis fails to be rejected for Internet use, professional use, and professional frequency, as statistical differences were not found for these variables.

# Recommendations

The results of this research project are potentially valuable to financial service providers, government entities, social media providers, academic researchers, and to the investing public in general. Although the survey population was a targeted audience that was not intended to be generalizable to a larger population, this study did produce statistically significant results specific to the target audience. Large-scale studies, such as the 2012 National Financial Capabilities Study produce valid and reliable data, but these large-scale studies are time-consuming and expensive to implement. Small-scale studies like this research project are useful for situations where external validity is not considered important to the specific research questions.

Specific to this research project, it would be recommended that outreach advertising for investment and financial planning information, placed on the LinkedIn platform be oriented towards college-educated men, within the age range of 35 to 74, and who have household income that exceeds $100,000 per year. This recommendation is based upon the demographics of the respondents.

The LinkedIn platform may not be as efficient for outreach to women. It is recommended that additional studies be conducted with the knowledge that women are less likely to use the internet for obtaining investment and financial planning information. The LinkedIn platform is also unlikely to be efficient for outreach to seniors over age 75. It is recommended that additional studies be conducted with the knowledge that the over-75 age group is not likely to use the internet for obtaining investment and financial planning information. It is also recommended that studies of the over-75 age group not be conducted via the internet, based upon the low response rate for this demographic.

It is a bit surprising that the LinkedIn platform may not be efficient for outreach to Millennials in the 18 to 34 age group, as the response rate for this group was similar to the low response rate for the over-75 age group. The Millennial generation, born digital, can be expected to learn differently than prior generations who learned to use computers later in adulthood. However, Millennials tend to be underinvested as they express distrust in financial institutions and the economy in general, even though they recognize that their money could be working harder. It is recommended that additional studies be conducted, perhaps through social media, to find efficient means for outreach to this demographic group.

The last recommendation is for additional studies to be conducted to build an understanding of how to outreach to the poorest income groups. Higher levels of financial literacy allow financial consumers to aspire to change their financial behavior. There is a positive association between financial literacy and financial satisfaction. There is also a positive association between financial satisfaction and quality of life. As the under $25,000 per year income group may not have adequate access to the internet, these studies would need to be conducted at the local neighborhood level.

## Limitations

It should be expected that the survey population from this study does not geographically represent the national population. As the survey was directed towards working and retired professionals that maintain profiles on LinkedIn, it should further be expected that the survey population does not reflect the national population either on an economic or an educational basis. There is no expectation that the survey reflects the population of LinkedIn users as a whole. As such, the survey population may be too uniform to provide statistically significant outcomes for the research question and hypotheses. In summary, it should be expected that the survey population will not be generalizable to a larger population.

Despite these limitations, this type of research has low implementation cost and is relatively easy to administer. Multiple targeted studies can be conducted in a timely manner. There are additional benefits in researching populations that are not generalizable to a large population. Targeted populations that better model an organization's outreach base can provide insight that is more useful to the sponsoring organization.

## Future Research

As financial literacy and education are important components of modern life, it is important to understand how financial consumers obtain information in this regard. There are a variety of media sources available to obtain investment and financial planning information, and demographic groups may differ in the use of media. More research is needed as to how financial consumers obtain investment and financial planning information. There needs to be greater understanding of where financial consumers source this information, and if the information is helpful in increasing financial literacy. It is also important to understand how different demographic groups make use of media to obtain investment and financial planning information. Improved understanding in this research area will provide needed insight on how to reach out to demographic groups that are currently underserved.

## Summary

It is recognized that socio-demographic factors have influence over the level of financial literacy for financial consumers. Key demographics for research include gender, age, and life situation demographics such as education-level and income-level. In this research study, which selected a population from the LinkedIn platform, statistical differences for gender, age, education-level, and income-level were confirmed. These differences help to support prior research in this field of study.

      Practical opportunities for commercial outreach to specific populations becomes evident through this type of research. Providers of investment and financial planning information can access their targeted audience more effectively by understanding the demographic profile of the audience, as well as the propensity of the demographic profile of the audience to respond. As this type of research is relatively easy to construct and administer, commercial outreach for providers of investment and financial planning information can be conducted in a cost-efficient and effective manner.

## Appendix A

## Survey Questionnaire

**Q1.  I find the internet to be very useful for obtaining investment and financial planning information.**

Strongly agree / Agree / Neither agree or disagree / Disagree / Strongly disagree

**Q2.  How frequently do you use the internet to search for and obtain investment and financial planning information?**

Daily / Weekly / Monthly / Quarterly / Annually / Never

**Q3.  I find consulting professionals (attorneys, CPAs, investment advisors, financial planners) to be very useful for obtaining investment and financial planning information.**

Strongly agree / Agree / Neither agree or disagree / Disagree / Strongly disagree

**Q4.  How frequently do you consult professionals to obtain investment and financial planning information?**

Daily / Weekly / Monthly / Quarterly / Annually / Never

**Q5.  I consider the BEST way to obtain investment and financial planning information to be:** _________________________________________________

___________________________________________________________________

___________________________________________________________________

**Q6.  I am confident that I am getting adequate information to make good investment and financial planning decisions.**

Strongly agree / Agree / Neither agree or disagree / Disagree / Strongly disagree

---

**The following information is being requested to assist with research study demographics:**

**Q7.  Do you identify as?**          female          or          male          or          other

**Q8.  Are you a college graduate?**          Yes          or          No

**Q9.  In what range is your age?**          18 to 34          35 to 54          55 to 74          75 and over

**Q10.  In what range is your household income?**          Under $25,000/year

$25,000/year to $100,000/year

Above $100,000/year